\documentclass[a4paper,11pt]{JHEP3} % 10pt is ignored!

\title{Unitary Quantum Physics with Time-Space Noncommutativity}

\author{A. P. Balachandran \\
	Physics Department, Syracuse University \\
        Syracuse, NY, 13244-1130, USA \\
	E-mail: \email{bal@phy.syr.edu}} 

\author{T. R. Govindarajan \\
        The Institute of Mathematical Sciences \\
        C. I. T. Campus Taramani, Chennai 600 113, India \\
	E-mail: \email{trg@imsc.res.in}} 

\author{C. Molina and P. Teotonio-Sobrinho \\
        Instituto de F\'{\i}sica, Universidade de S\~{a}o Paulo \\
        C.P. 66318, S\~{a}o Paulo, SP, 05315-970, Brazil \\
	E-mail: \email{cmolina@fma.if.usp.br}, \email{teotonio@fma.if.usp.br}
        }

\preprint{IMSc/2004/06/24,SU-4252-795,USP-IF-1593/2004}

\abstract{In this work quantum physics in noncommutative spacetime is
developed. It is based on the work of Doplicher et al. which allows
for time-space noncommutativity. The Moyal plane is treated in detail.
In the context of noncommutative quantum mechanics, some important
points are explored, such as the formal construction of the theory,
symmetries, causality, simultaneity and observables. The dynamics
generated by a noncommutative  Schr\"{o}dinger equation is studied. We
prove in particular the following: suppose the Hamiltonian $H$ of a
quantum mechanical particle on spacetime $\mathbb{R}^{N-1} \times
\mathbb{R}$ has no explicit time dependence, and the spatial
coordinates commute in its noncommutative form (the only
noncommutativity being between time and a space coordinate). Then the
noncommutative version $\hat{H}$ of $H$ and $H$ have identical
spectra.}


\begin{document}

%%%%%%%%%%%%%%%%%%%%%%%%%%%%%%%%%%%%%%%%%%%%%%%%%%%%%%%%%%%%%%%%%%%%%%%%%%%%%
\section{Introduction}
%%%%%%%%%%%%%%%%%%%%%%%%%%%%%%%%%%%%%%%%%%%%%%%%%%%%%%%%%%%%%%%%%%%%%%%%%%%%%

Considerations based on quantum gravity and black hole physics led
to the suggestion several years ago \cite{Doplicher} that spacetime
commutativity may be lost at the smallest scale, the commutators of
time and space coordinates ($\hat{x}_{0}$ and $\hat{x}_{i}$) having
the form 
\begin{equation}
\left[\hat{x}_{\mu},\hat{x}_{\nu}\right]=i\theta_{\mu\nu}\mathbb{I}
\,\, ,
\label{nc_relation}
\end{equation}
with $\theta_{\mu\nu}$ being constants of the order of the square
of Planck length. String theory also incorporates relations like
(\ref{nc_relation}).  

Commutators such as (\ref{nc_relation}) actually have a much more
ancient origin. They seem to have first appeared in a letter from
Heisenberg to Peierls in 1930 \cite{history}. Spacetime
noncommutativity was later revived by Snyder \cite{Snyder} who sought
to use it to regularize quantum field theories (qft's), and then by
Yang \cite{Yang}. Madore \cite{Madore} also attributes similar ideas
to Dirac. Among the early works in noncommutative spacetime is that of Kempf et
al. \cite{Kempf}. A subsequent related work is that of Lizzi et
al. \cite{Lizzi}.   

Conventional studies of (\ref{nc_relation}) assume that $\theta_{0i}=0$
so that the time coordinate commutes with the rest. There are even
claims that qft's based on (\ref{nc_relation}) are nonunitary if
$\theta_{0i}\ne0$.   

In contrast, in a series of fundamental papers, Doplicher et
al. \cite{Doplicher} have studied (\ref{nc_relation}) in complete
generality, without assuming that $\theta_{0i}\ne0$ and developed
unitary qft's which are ultraviolet finite to all orders. 

This paper is based on the work of Doplicher et al. Using their ideas,
we systematically develop unitary quantum mechanics based on
(\ref{nc_relation}). It indicates where to look for phenomenological
consequences of (\ref{nc_relation}) and also easily leads to the
considerations of Doplicher  et al. \cite{Doplicher} on qft's. 

The relation (\ref{nc_relation})  will be treated with  $\theta$ being
constant. Our focus is on time and its noncommutativity with spatial
coordinates. For this purpose, it is enough to examine
(\ref{nc_relation}) on a $(1+1)$-spacetime and replace it by  
\begin{equation}
\left[\hat{x}_{\mu},\hat{x}_{\nu}\right] = i\theta
\varepsilon_{\mu\nu} \mathbb{I}   \,\, , 
\varepsilon_{\mu\nu} = - \varepsilon_{\nu\mu} \,\, , 
\varepsilon_{01} = 1 \,\, .
\end{equation}
We assume with no loss of generality that $\theta>0$, as we can
change its sign by flipping $\hat{x}_{1}$ to $-\hat{x}_{1}$. We denote
by $\mathcal{A}_{\theta}\left(\mathbb{R}^{2}\right)$ the unital algebra
generated by $\hat{x}_{0}$, $\hat{x}_{1}$ and $\mathbb{I}$.

%%%%%%%%%%%%%%%%%%%%%%%%%%%%%%%%%%%%%%%%%%%%%%%%%%%%%%%%%%%%%%%%%%%%%%%%%%%%%
\section{Qualitative Remarks}
%%%%%%%%%%%%%%%%%%%%%%%%%%%%%%%%%%%%%%%%%%%%%%%%%%%%%%%%%%%%%%%%%%%%%%%%%%%%%

\subsection{Symmetries}

If a group of transformations cannot be implemented on the algebra
$\mathcal{A}_{\theta}\left(\mathbb{R}^{2}\right)$ generated by $\hat{x}_{\mu}$
with relation (\ref{nc_relation}), then it is not likely to be
a symmetry of any physical system based on (\ref{nc_relation})
\cite{note}. So let us check what are the automorphisms of
(\ref{nc_relation}).

\subsubsection{Translations}

First we readily see that spacetime translations $\mathcal{U}(\vec{a})$,
$\vec{a}=(a_{0},a_{1})$, $a_{\mu}\in\mathbb{R}$, are automorphisms
of $\mathcal{A}_{\theta}\left(\mathbb{R}^{2}\right)$: with 
\begin{equation}
\mathcal{U}(\vec{a})\hat{x}_{\mu}=\hat{x}_{\mu}+a_{\mu} \,\, ,
\end{equation}
we see that 
\begin{equation}
\left[ \mathcal{U}(\vec{a}) \hat{x}_{\mu} , \mathcal{U}(\vec{a})
\hat{x}_{\nu}\right] =
i\theta\varepsilon_{\mu\nu} \,\,.
\end{equation}
The existence of these automorphisms allows the possibility of energy-momentum
conservation. The time-translation automorphism 
\begin{equation}
U(\tau) := \mathcal{U} \left((\tau,0)\right)
\label{time_translation}
\end{equation}
is of particular importance. Without it, we cannot formulate conventional
quantum physics. 

The infinitesimal generators of $\mathcal{U}(\vec{a})$ can be defined
by writing  
\begin{equation}
\mathcal{U}(\vec{a}) = e^{- ia_{0}\hat{P}_{0} + ia_{1}\hat{P}_{1}} \,\, .
\end{equation}
Then we have 
\begin{equation}
\hat{P}_{0} = -\frac{1}{\theta}\textrm{ad}\,\hat{x}_{1} \,\, , \,\,
\hat{P}_{1} = -\frac{1}{\theta}\textrm{ad}\,\hat{x}_{0} \,\, , \,\,
\textrm{ad} \, \hat{x}_{\mu}\hat{a} \equiv
\left[\hat{x}_{\mu},\hat{a}\right] \,\, ,\,\,
\hat{a}\in\mathcal{A}_{\theta}\left(\mathbb{R}^{2}\right) \,\, .
\label{generators}
\end{equation}
The relations (\ref{generators}) show that the automorphisms
$\mathcal{U}(\vec{a})$ are inner.

\subsubsection{The Lorentz Group}

It is a special feature of two dimensions that the $(2+1)$ connected
Lorentz group is an inner automorphism group of (\ref{nc_relation}).
The above group is the two-dimensional projective symplectic group,
the symplectic group quotiented by its center $\mathbb{Z}_{2}$. Its
generators are ${\textrm{ad}\hat{J}_{3}}$ and ${\textrm{ad}\hat{K}_{a}}$,
where 
\begin{equation}
\hat{J}_{3} =
\frac{1}{4\theta}\left(\hat{x}_{0}^{2}+\hat{x}_{1}^{2}\right) \,\, ,\,\,
\hat{K}_{1} = 
\frac{1}{4\theta}\left(\hat{x}_{0}\hat{x}_{1}+\hat{x}_{1}\hat{x}_{0}\right)
\,\, ,\,\,
\hat{K}_{2} = 
\frac{1}{4\theta}\left(\hat{x}_{0}^{2}-\hat{x}_{1}^{2}\right) \,\, ,
\end{equation}
with the ad notation explained by (\ref{generators}).
Although this group generates inner automorphisms, it cannot be
implemented on the quantum Hilbert space because, as we shall later
see, $\hat{x}_{0}$ is not an operator on the physical Hilbert space.  

The algebra $\mathcal{A}_{\theta}\left(\mathbb{R}^{2}\right)$ is
a {*}-algebra with  
\begin{equation}
\hat{x}_{\mu}^{*}=\hat{x}_{\mu} \,\, .
\end{equation}
We note that 
\begin{equation}
\hat{J}_{3}^{*} = \hat{J}_{3} \,\, , \,\,
\hat{K}_{a}^{*} = \hat{K}_{a} \,\, .
\end{equation}

\subsubsection{P, T, C Symmetries}

There are certain important transformations which are automorphisms
for $\theta=0$, but not for $\theta\ne0$. One such is parity $P$:
\begin{equation}
P:\hat{x}_{0} \rightarrow\hat{x}_{0} \,\, , \,\,
  \hat{x}_{1} \rightarrow-\hat{x}_{1} \,\, , \,\,
  \mathbb{I} \rightarrow\mathbb{I} \,\, .
\end{equation}
We want it furthermore to be linear. But that does not preserve
(\ref{nc_relation}) if $\theta\ne0$: 
\begin{equation}
P: \left[\hat{x}_{0},\hat{x}_{1}\right] \rightarrow-
   \left[\hat{x}_{0},\hat{x}_{1}\right] \,\, , \,\, 
   i\theta\mathbb{I}\rightarrow i\theta\mathbb{I}\,\, .
\end{equation}

In contrast, time-reversal $T$, 
\begin{equation}
T: \hat{x}_{0} \rightarrow- \hat{x}_{0} \,\, , \,\,
   \hat{x}_{1} \rightarrow \hat{x}_{1}
\end{equation}
is anti-linear, 
\begin{equation}
T: i\theta\mathbb{I}\rightarrow-i\theta\mathbb{I} \,\, ,
\end{equation}
so that it is an automorphism of
$\mathcal{A}_{\theta}\left(\mathbb{R}^{2}\right)$. 

Hence any theory based on (\ref{nc_relation}) violates $P$ and $PT$.
Superficially there seems to be no problem in writing charge conjugation
invariant models based on (\ref{nc_relation}). For such models, $CPT$
will also fail to be a symmetry \cite{CPT_violation}. 

The symmetries $P$ and $PT$ are automorphisms of the algebra which
is the direct sum of $\mathcal{A}_{\theta}\left(\mathbb{R}^{2}\right)$
and $\mathcal{A}_{-\theta}\left(\mathbb{R}^{2}\right)$. In that case,
spacetime will have two leaves. The Doplicher et al. models are based
on such algebras.

\subsubsection{Further Automorphisms}

As fully discussed in \cite{Susskind,Jackiw}, infinitesimal
transformations 
$\hat{x}_{\mu}\rightarrow \hat{x}_{\mu}+\delta \hat{x}_{\mu}$ 
of the form 
$\delta \hat{x}_{\mu} = \hat{f}_{\mu}(\hat{x}_{0},\hat{x}_{1})$
generate automorphisms of
$\mathcal{A}_{\theta}\left(\mathbb{R}^{2}\right)$ if the condition
\begin{equation}
[\hat{f}_{\mu},\hat{x}_{\nu}] + [\hat{x}_{\mu},\hat{f}_{\nu}]=0
\end{equation}
is satisfied. The associated group of transformations exhausts  the
noncommutative version of the area-preserving transformations (in two
dimensions and connected to the identity), and includes the Lorentz
group as a particular case.    

\subsection{Causality}

It is impossible to localize (the representation of) ``coordinate''
time $\hat{x}_{0}$ in $\mathcal{A}_{\theta}\left(\mathbb{R}^{2}\right)$
sharply. Any state will have a spread in the spectrum of
$\hat{x}_{0}$. This leads to failure of causality as explained by
Chaichian et al. \cite{Chaichian}. 

The following important point was emphasised to us by Doplicher
\cite{Doplicher_private}. In quantum mechanics, if $\hat{p}$ is
momentum, $\exp(i\xi\hat{p})$ is spatial translation by amount
$\xi$. This $\xi$ is not the eigenvalue of the position operator
$\hat{x}$. In the same way, the amount $\tau$ of time translation in
(\ref{time_translation}) is not ``coordinate time'', the eigenvalue of
$\hat{x}_{0}$ \cite{Doplicher}. It makes sense to talk about a state
and its translate by $U(\tau)$. For $\theta=0$, it is possible to
identify coordinate time with $\tau$: the former is just a parameter
we need for labelling time-slices of spacetime and increasing with
$\tau$. But for $\theta\ne0$, $\hat{x}_{0}$ is an operator not
commuting with $\hat{x}_{1}$, and cannot be interchanged with $\tau$.  

Concepts like duration of an experiment for $\theta=0$ \cite{Brunetti}
are expressed using $U(\tau)$. They carry over to the noncommutative
case too.

\subsection{The Spin-Statistics Connection}

With loss of causality, one loses local qft's as well. As the best
proofs of the spin-statistics connection require locality
\cite{spin-statistics}, we can anticipate the breakdown of this
connection as well when
$\mathcal{A}_{\theta}\left(\mathbb{R}^{2}\right)$ is generalised to
$(3+1)$ dimensions. Precision experiments to test the spin-statistics
connection are possible \cite{Capri_meeting}. If signals for this
violation due to $\theta\ne0$ can be derived, good phenomenological
bounds on $|\theta|$ should be possible.

%%%%%%%%%%%%%%%%%%%%%%%%%%%%%%%%%%%%%%%%%%%%%%%%%%%%%%%%%%%%%%%%%%%%%%%%%%%%%
\section{Representation Theory }
%%%%%%%%%%%%%%%%%%%%%%%%%%%%%%%%%%%%%%%%%%%%%%%%%%%%%%%%%%%%%%%%%%%%%%%%%%%%%

Observables, states and dynamics of quantum theory are to be based
on the algebra $\mathcal{A}_{\theta}\left(\mathbb{R}^{2}\right)$
defined by (\ref{nc_relation}). The formalism for their construction,
using the methods of the GNS approach \cite{Haag} in the commutative and
non-commutative contexts, will be explored in the following.  

Now to each $\hat{\alpha}\in\mathcal{A}_{\theta}\left(\mathbb{R}^{2}\right)$,
we can canonically associate its left and right regular representations
$\hat{\alpha}^{L}$ and $\hat{\alpha}^{R}$,   
\begin{equation}
\hat{\alpha}^{L}\hat{\beta} = \hat{\alpha}\hat{\beta} \,\, , \,\,
\hat{\alpha}^{R}\hat{\beta} = \hat{\beta}\hat{\alpha} \,\, , \,\,
\hat{\beta}\in\mathcal{A}_{\theta}\left(\mathbb{R}^{2}\right) \,\, ,
\label{Left_Right_reps}
\end{equation}
with $\hat{\alpha}^{L}\hat{\beta}^{L}=\left(\hat{\alpha}\hat{\beta}\right)^{L}$
and $\hat{\alpha}^{R}\hat{\beta}^{R}=\left(\hat{\beta}\hat{\alpha}\right)^{R}$.
The carrier space of this representation is
$\mathcal{A}_{\theta}\left(\mathbb{R}^{2}\right)$ itself.

But such representations are not enough for quantum physics. An ``inner''
product on $\mathcal{A}_{\theta}\left(\mathbb{R}^{2}\right)$ is
needed for an eventual construction of a Hilbert space.  

Doplicher et al. get this inner product using positive maps. 
Consider a map
\mbox{$\chi: \mathcal{A}_{\theta} \left(\mathbb{R}^{2}\right)
\rightarrow \mathbb{C}$} with the usual properties of
$\mathbb{C}$-linearity and preservation of $*$:
$\chi\left(\hat{\alpha}^{*}\right)=\overline{\chi(\hat{\alpha})}$ (bar
meaning complex conjugation). It is a positive map if  
\begin{equation}
\chi\left(\hat{\alpha}^{*}\hat{\alpha}\right) \ge 0 \,\, .
\end{equation}

Given such a map, we can set 
$\left\langle \hat{\alpha},\hat{\beta}\right\rangle
= \chi\left(\hat{\alpha}^{*}\hat{\beta}\right)$.  
It will be a scalar product if $\chi\left(\hat{\alpha}^{*}\alpha\right)=0$
implies $\hat{\alpha}=0$. If that is not the case, it is necessary to
eliminate nonzero vectors of zero norm (null vectors).    

We illustrate these ideas first in the context of the commutative
case, when $\theta=0$. Then we generalise these ideas to (\ref{nc_relation})
and in particular we discuss two positive maps. The first, due to
Doplicher et al., is based on a symbol of the operators.
The second uses the Voros symbol based on coherent states. After some
analysis, we show that they lead to identical physics.

\subsection{The Commutative Case}

\subsubsection{The Positive Map}

The algebra $\mathcal{C}$ in the commutative case is  
$\mathcal{A}_{0}\left(\mathbb{R}^{2}\right) =
C^{\infty}\left(\mathbb{R}\times\mathbb{R}\right)$,  
the product being point-wise multiplication, and $*$ being complex
conjugation. If $\psi\in\mathcal{C}$, then
$\psi(x_{0},x_{1})\in\mathbb{C}$, where $(x_{0},x_{1})$ are
coordinates of $\mathbb{R}^{2}$.

There is no distinction now between $\hat{\alpha}^{L}$ and $\hat{\alpha}^{R}$:
$\hat{\alpha}^{L}=\hat{\alpha}^{R}$.   

There is actually a family of
positive maps $\chi_{t}$ of interest obtained by integrating $\psi$
in $x_{1}$ at ``time'' $t$: 
\[
\chi_{t}(\psi) = \int\, dx_{1}\,\psi(t,x_{1}) \,\, ,
\]
\begin{equation}
\chi_{t}(\psi^{*}\psi) \ge 0 \,\, .
\end{equation}
This defines a family of spaces $\mathcal{C}_{t}$ with a positive-definite
sesquilinear form (an ``inner product'') $(.\,,\,.)_{t}$:  
\begin{equation}
\left(\psi,\varphi \right)_{t} = \int \, dx_{1} \, 
\psi^{*}(t,x_{1}) \varphi (t,x_{1}) \,\, . 
\label{commutative_product}
\end{equation}
(We associate $\chi_{t}$ with $\mathcal{C}$ to get $\mathcal{C}_{t}$.)

\subsubsection{The Null Space $\mathcal{N}_{t}^{0}$}

Every function $\hat{\alpha}$ which vanishes at time $t$ is a two-sided
ideal $\mathcal{I}_{t}^{\theta=0}:=\mathcal{I}_{t}^{0}$ of $\mathcal{C}$.
As elements of $\mathcal{C}_{t}$ , they become null vectors
$\mathcal{N}_{t}^{0}$ in the inner product
(\ref{commutative_product}). (We associate $\chi_{t}$ also to
$\mathcal{I}_{t}^{0}$ to get $\mathcal{N}_{t}^{0}$.) Thus as in the
GNS construction \cite{Haag}, we can quotient by these vectors and work with
$\mathcal{C}_{t}/\mathcal{N}_{t}^{0}$. For elements
$\psi+\mathcal{N}_{t}^{0}$ and $\chi+\mathcal{N}_{t}^{0}$ in
$\mathcal{C}_{t}/\mathcal{N}_{t}^{0}$, the scalar product is 
\begin{equation}
\left(\psi+\mathcal{N}_{t}^{0} , \chi+\mathcal{N}_{t}^{0}\right)_{t} =
\left(\psi,\chi\right)_{t} \,\, .
\label{null-space-product}
\end{equation}
There are no non-trivial vectors of zero norm now. The completion
$\overline{\mathcal{C}_{t}/\mathcal{N}_{t}^{0}}$ of
$\mathcal{C}_{t}/\mathcal{N}_{t}^{0}$ in this scalar product gives a
Hilbert space $\widehat{\mathcal{H}}_{t}^{0}$. We have also that
$\mathcal{C}_{t}/\mathcal{I}_{t}^{0}$ acts on it faithfully,
preserving its $*$, 
\begin{equation}
\left(\psi+\mathcal{I}_{t}^{0}\right)^{*} =
\psi^{*}+\left(\mathcal{I}_{t}^{0}\right)^{*} =
\psi^{*}+\mathcal{I}_{t}^{0} \,\, .
\end{equation}
In the expression above, $S^{*}$ is the set obtained from $S$ by taking
the complex conjugate of each element. Hence
$\left(\mathcal{I}_{t}^{0}\right)^{*}=\mathcal{I}_{t}^{0}$.

\subsubsection{The Quantum Mechanical Hilbert Space $\mathcal{H}_{t}^{0}$}

The quantum mechanical Hilbert space however is not
$\widehat{\mathcal{H}}_{t}^{0}$. It is constructed in a different way,
starting from a subspace
$\tilde{\mathcal{H}}_{0,t}\subset\mathcal{C}_{t}$ which contains only
$\left\{ 0\right\} $ as the null vector:  
\begin{equation}
\tilde{\mathcal{H}}_{0,t}\cap\mathcal{N}_{t}^{0}=\left\{ 0\right\} \,\, .
\end{equation}
(The subscript $0$ on $\tilde{\mathcal{H}}_{0,t}$ denotes the value of
$\theta$.) Then $\chi_{t}$ is a good scalar product on
$\tilde{\mathcal{H}}_{0,t}$ and the quantum mechanical Hilbert space
is given by
$\mathcal{H}_{t}^{0}=\overline{\tilde{\mathcal{H}}_{0,t}}$, the
completion of $\tilde{\mathcal{H}}_{0,t}$.    

The subspace $\tilde{\mathcal{H}}_{0,t}$ depends on the Hamiltonian
$H$ and is chosen as follows. Suppose first that $H$ is a
time-independent Hamiltonian on commutative spacetime, self-adjoint on
the standard quantum mechanical Hilbert space
$L^{2}\left(\mathbb{R}\right)$. It acts on $\mathcal{C}_{t}$ and obeys 
\begin{equation}
\left(\psi,H\chi\right)_{t}=\left(H\psi,\chi\right)_{t} \,\, .
\end{equation}

We now pick the subspace $\tilde{\mathcal{H}}_{0,t}$ of
$\mathcal{C}_{t}$ by requiring that vectors in $\mathcal{C}_{t}$ obey
the time-dependent Schr\"{o}dinger equation: 
\begin{equation}
\tilde{\mathcal{H}}_{0,t} = \left\{ \psi\in\mathcal{C}_{t}:
\left(i\partial_{x_{0}}-H\right)\psi(x_{0},x_{1}) = 0 \right\} \,\,.
\end{equation}
The operator $i\partial_{x_{0}}$ is not ``hermitian'' on all
vectors of $\mathcal{C}_{t}$: 
\begin{equation}
\left(\psi,i\partial_{x_{0}}\chi\right)_{t} \ne
\left(i\partial_{x_{0}}\psi, \chi\right)_{t} \,\,
\textrm{for generic} \,\,
\psi,\chi\in\mathcal{C}_{t}\,\, ,
\end{equation}
but on $\tilde{\mathcal{H}}_{0,t}$, it equals $H$ and does fulfill
this property: 
\begin{equation}
\left(\psi,i\partial_{x_{0}}\chi\right)_{t} = \left(i\partial_{x_{0}}
  \psi,\chi\right)_{t} \,\,
\textrm{for generic}\,\,
\psi,\chi\in\tilde{\mathcal{H}}_{0,t} \,\, .
\end{equation}
Since $\left[i\partial_{x_{0}},H\right]=0$, both $i\partial_{x_{0}}$
and $H$ leave the subspace $\tilde{\mathcal{H}}_{0,t}$ invariant: 
\begin{equation}
i\partial_{x_{0}}\tilde{\mathcal{H}}_{0,t} =
H\tilde{\mathcal{H}}_{0,t} \subseteq \tilde{\mathcal{H}}_{0,t} \,\, . 
\end{equation}

We see also that since  
\begin{equation}
\psi(x_{0}+\tau,x_{1}) = 
\left(e^{-i\tau\left(i\partial_{x_{0}}\right)}\psi\right)(x_{0},x_{1})=
\left(e^{-i\tau H}\psi\right)(x_{0},x_{1}) \,\, ,
\end{equation}
time evolution preserves the norm of
$\psi\in\tilde{\mathcal{H}}_{0,t}$. Therefore if it vanishes at
$x_{0}=t$, it vanishes identically and is the zero element of
$\tilde{\mathcal{H}}_{0,t}$: the only null vector in
$\tilde{\mathcal{H}}_{0,t}$ is $0$:  
\begin{equation}
\mathcal{N}_{t}^{0}\cap\tilde{\mathcal{H}}_{0,t}=\left\{ 0\right\} \,\, .
\end{equation}
That means that $\chi_{t}$ gives a true scalar product on
$\tilde{\mathcal{H}}_{0,t}$. The completion of
$\tilde{\mathcal{H}}_{0,t}$ is the quantum Hilbert space
$\mathcal{H}_{t}^{0}$.   

We can find no convenient inclusion of $\mathcal{H}_{t}^{0}$ in
$\widehat{\mathcal{H}}_{t}^{0}$. The reason is that
$\mathcal{N}_{t}^{0}$ is not in the kernel of
$\left(i\partial_{x_{0}}-H\right)$, only its zero vector is.   

Elements of $\tilde{\mathcal{H}}_{0,t}$ are very conventional. Let
$\hat{x}_{\mu}$ be coordinate functions
($\hat{x}_{\mu}(x_{0},x_{1})=x_{\mu}$) so that
$i\partial_{x_{0}}\hat{x}_{\mu}=i\delta_{0\mu}$, and let $\psi_{0}$ be
a constant function of $x_{0}$ so that $i\partial_{x_{0}}\psi_{0}=0$. 
Then 
\begin{equation}
\psi = e^{-i\hat{x}_{0}H}\psi_{0}\in\tilde{\mathcal{H}}_{0,t} \,\, .
\end{equation}
Under time evolution by amount $\tau$, $\psi$ becomes 
\begin{equation}
e^{-i\tau H}\psi =
e^{-i\left(\hat{x}_{0}+\tau\right)H}\psi_{0}\in\tilde{\mathcal{H}}_{0,t} \,\, .
\end{equation}
The conceptual difference between coordinate time $\hat{x}_{0}$
and amount of time translation $\tau$ is apparent here. As one learns
from Doplicher et al. \cite{Doplicher}, this difference cannot be
ignored with spacetime noncommutativity. 

As $\psi_{0}$ is constant in $x_{0}$, its values may be written as
$\psi_{0}(x_{1})$.

\subsubsection{On Observables}

An observable $\hat{K}$ has to respect the Schr\"{o}dinger constraint
and leave $\tilde{\mathcal{H}}_{0,t}$ (and hence $\mathcal{H}_{t}^{0}$)
invariant. This means that 
\begin{equation}
\left[i\partial_{x_{0}}-H,\hat{K}\right] = 0 \,\, .
\label{Schrodinger_constraint}
\end{equation}
 
Let $\hat{L}$ be any operator with no explicit time dependence so that
$\hat{L}$ is a function of $\hat{x}_{1}$ and momentum. Then 
\begin{equation}
\hat{K} = e^{-i\hat{x}_{0}H}\hat{L}e^{+i\hat{x}_{0}H} 
\label{hat-K}
\end{equation}
is an observable. We have also that $\hat{K}$ acts on $\psi$ in a
familiar manner:  
\begin{equation}
\hat{K}\psi = \left(\hat{L}\psi_{0}\right)e^{-i\hat{x}_{0}H} \,\, .
\end{equation}
Under time translation, $\hat{x}_{0}$ in $\hat{K}$ shifts to
$\hat{x}_{0}+\tau$ as it should:
\begin{equation}
e^{-i\tau H} \hat{K} e^{+i\tau H} =
e^{-i\left(\hat{x}_{0} + \tau\right)H} \hat{L} e^{+i\left(\hat{x}_{0}
  + \tau\right)H}\,\,.  
\end{equation}

Response under time-translations is dynamics, it gives
time-evolution. Just as in the conventional approach, here and
elsewhere we should time-evolve either vector states (Schr\"{o}dinger
representation) or observables (Heisenberg representation). One can
also formulate the interaction representation. 

A final important point is the following. The observables have the
expected reality properties. In particular, $\mathcal{C}$ is a $*$-algebra,
with star being complex conjugation, denoted here by a bar. So are
the functions $\hat{L}$ on $\mathbb{R}^2$ which are constant in
$x_{0}$, that is, functions of position only. If $\hat{K}$ is its
image on $\mathcal{H}_{t}^{0}$, as in (\ref{hat-K}), then $\overline{\hat{L}}$
has image $\hat{K}^{\dagger}$: we have a $*$-representation of these
functions. Momentum too is a self-adjoint operator on
$\mathcal{H}_{t}^{0}$.

\subsubsection{Time-dependent $H$}

Next suppose that $H$ has $\hat{x}_{0}$-dependence: 
\begin{equation}
\left[i\partial_{x_{0}},H\right] \ne 0 \,\, .
\end{equation}
This means that $H$ is a function of $\hat{x}_{0}$ and other operators
like $\hat{x}_{1}$ and momentum, and we should write for the Hamiltonian
$H\left(\hat{x}_{0},\hat{x}_{1},-i\partial_{x_{1}}\right)$. There
is no factor-ordering problem involving $\hat{x}_{0}$ here. We can
substitute a real variable $x_{0}$ for $\hat{x}_{0}$ and get the
operator $H\left(x_{0},\hat{x}_{1},-i\partial_{x_{1}},\ldots\right)$
without ambiguity.  

The Schr\"{o}dinger constraint (\ref{Schrodinger_constraint}) remains
intact, but $\psi\in\tilde{\mathcal{H}}_{0,t}$ has a different expression: 
\[
\psi = U\left(\hat{x}_{0},\tau_{I}\right)\psi_{0} \,\, ,
\]
\begin{equation}
U\left(\hat{x}_{0},\tau_{I}\right) =
\left.T\,\exp\left[-i\int_{\tau_{I}}^{x_{0}}\, dx'_{0}\,
H \left(x'_{0},\hat{x}_{1},-i\partial_{x_{1}},\ldots\right)\right]
\right|_{x_{0}=\hat{x}_{0}} \,\, .
\end{equation}
where $\tau_{I}$ is the initial time at which $\psi=\psi_{0}$ (which
depends only on $\hat{x}_{1}$), and $T$ is time ordering in $x'_{0}$.

Time translation by amount $\tau$ shifts $\hat{x}_{0}$ to $\hat{x}_{0}+\tau$
in $U$ as before: $U\left(\hat{x}_{0},\tau_{I}\right)\rightarrow
U\left(\hat{x}_{0}+\tau,\tau_{I}\right)$. Observables are constructed
from $\hat{L}$ using $U$ and have familiar properties.

\subsubsection{Is Time an Observable?}

What we have described above leads to conventional physics. Just as
in the latter, here too, $\hat{x}_{0}$ is not an observable as it
does not commute with $i\partial_{x_{0}}-H$: 
\begin{equation}
\left[\hat{x}_{0},i\partial_{x_{0}}-H\right] = -i\mathbb{I} \,\, .
\end{equation}
Transformations with $\exp\left(-i\hat{x}_{0}H\right)$ or $U$ does
not affect $\hat{x}_{0}$. So we cannot construct an observable therefrom
as we did to get $\hat{K}$ from $\hat{L}$.

\subsubsection{On the Time-dependence of $\mathcal{H}_{t}^{0}$}

In conventional quantum physics, the Hilbert space has no time-dependence,
whereas $\mathcal{H}_{t}^{0}$ has a label $t$. This is puzzling.

But the puzzle is easy to resolve: $\mathcal{H}_{t}^{0}$ is independent
of $t$. Thus the solutions $\psi$ of the Schr\"{o}dinger constraint
do not depend on $t$ and are elements of every $\mathcal{H}_{t}^{0}$.
Their scalar products too are independent of $t$ because of the unitarity
of $H$. There is thus only one Hilbert space which we call $\mathcal{H}_{0}$
($0$ standing for the value of $\theta$). We also denote
$\tilde{\mathcal{H}}_{0,t}$ by $\tilde{\mathcal{H}}_{0}$
henceforth. Further the observables have no explicit $t$-dependence
and act on $\mathcal{H}_{0}$ as in standard quantum theory.

\subsection{The Noncommutative Case}

The above discussion shows that for quantum theory, what we need are:
(1) a suitable inner product on 
$\mathcal{A}_{\theta}\left(\mathbb{R}^{2}\right)$;
(2) a Schr\"{o}dinger constraint on
$\mathcal{A}_{\theta}\left(\mathbb{R}^{2}\right)$;
and (3) a Hamiltonian $\hat{H}$ and observables which act on the
constrained subspace of $\mathcal{A}_{\theta}\left(\mathbb{R}^{2}\right)$.
We also require that (1) is compatible with the self-adjointness of
$\hat{H}$ and classically real observables. 

We now consider these items one by one.

\subsubsection{The Inner Product}

There are several suitable inner products at first sight. But we shall
later argue that they are all equivalent.

The first inner product is based on symbol calculus. If
$\hat{\alpha}\in\mathcal{A}_{\theta}\left(\mathbb{R}^{2}\right)$, 
we write it as
\begin{equation}
\hat{\alpha} = \int d^{2}k \,
\tilde{\alpha}(k) e^{ik_{1}\hat{x}_{1}}e^{ik_{0}\hat{x}_{0}} \,\, ,
\label{alpha}
\end{equation}
and associate the symbol $\alpha_{S}$ with $\hat{\alpha}$ where
\begin{equation}
\alpha_{S}(x_{0},x_{1}) = \int d^{2}k \,
\tilde{\alpha}(k) e^{ik_{1}x_{1}}e^{ik_{0}x_{0}} \,\, .
\label{alpha-s-symbol}
\end{equation}
The symbol is a function on $\mathbb{R}^{2}$. It is not the Moyal
symbol. For the latter, the exponentials in (\ref{alpha}) must be
written as
$\exp \left(ik_{1}\hat{x}_{1}+ik_{0}\hat{x}_{0}\right)$. 

Using this symbol, we can define a positive map $S_{t}$ by
\begin{equation}
S_{t}\left(\hat{\alpha}\right) = \int dx_{1} \, \alpha_{S}(t,x_{1})
\,\, . 
\label{S-positive-map}
\end{equation}

Properties of $S_{t}$ are similar to $\chi_{t}$. In particular it
gives the inner product $\left(.,.\right)_{t}$, where
\begin{equation}
\left(\hat{\alpha},\hat{\beta}\right){}_{S_t} =
S_{t}\left(\hat{\alpha}^{*}\hat{\beta}\right)=
\int dx_{1} \, \alpha_{S}^{*}(t,x_{1})\beta_{S}(t,x_{1}) \,\, .
\label{S-inner-product}
\end{equation}
This inner product has null vectors
$\mathcal{N}_{t}^{\theta}:\hat{\alpha}\in\mathcal{N}_{t}^{\theta}$ if
$\alpha_{S}(t,.)=0$. But we will not consider them further as the
physical Hilbert space $\mathcal{H}_{t}^{\theta}$ is not obtained from
$\mathcal{A}_{\theta}\left(\mathbb{R}^{2}\right)/\mathcal{N}_{t}^{\theta}$.  

A second inner product can be constructed using the Voros
map, based on the coherent states associated with
(\ref{nc_relation}). 
Let
\begin{equation}
a = \frac{\hat{x}_{0}+i\hat{x}_{1}}{\sqrt{2\theta}} \,\, , \,\, 
a^{\dagger} = \frac{\hat{x}_{0}-i\hat{x}_{1}}{\sqrt{2\theta}}\,\,,\,\,
\left[a,a^{\dagger}\right] = \mathbb{I} \,\, ,
\end{equation}
and introduce the coherent states 
\begin{equation}
\left|z = x_{0} + ix_{1} \right\rangle =
e^{\frac{1}{\sqrt{2\theta}} \left(za^{\dagger} - \bar{z}a\right)}
\left|0 \right\rangle \,\,.
\end{equation}
The Voros or coherent state symbol of an operator
$\hat{\alpha}\in\mathcal{A}_{\theta}\left(\mathbb{R}^{2}\right)$ is
the function $\alpha_{V}$ on $\mathbb{R}^{2}$ where  
\begin{equation}
\alpha_{V}\left(x_{0},x_{1}\right) = 
\left\langle z\right|\hat{\alpha}\left|z\right\rangle \,\, .
\label{Vorus_map}
\end{equation}
The positive map $V_{t}$ is then defined by 
\begin{equation}
V_{t}\left(\hat{\alpha}\right) = \int dx_{1} \,
\alpha_{V}\left(t,x_{1}\right) \,\, .
\label{Vorus_positive_map}
\end{equation}  

As the symbol of a positive operator $\hat{\alpha}^{*}\hat{\alpha}$
is a non-negative function, the positivity of the map $V_{t}$ is
manifest from (\ref{Vorus_positive_map}). There are also no nontrivial
null vectors in the scalar product 
\begin{equation}
\left(\hat{\alpha},\hat{\beta}\right)_{V_t} =
V_{t}\left(\hat{\alpha}^{*}\hat{\beta}\right)
\label{Voros_scalar_product}
\end{equation}
as one can show. But that result is not important for what follows
as the Hilbert space is obtained only after constraining the vector
states by the noncommutative Schr\"{o}dinger equation.

\subsubsection{The Schr\"{o}dinger Constraint}

The noncommutative analogue ``$i\frac{\partial}{\partial x_{0}}$''
of the corresponding commutative operator is 
\begin{equation}
i\frac{\partial}{\partial x_{0}}\equiv\hat{P}_{0} =
-\frac{1}{\theta}{\rm ad}\,\hat{x}_{1} \,\, ,
\end{equation}
since 
\begin{equation}
-\frac{1}{\theta}{\rm
 ad}\,\hat{x}_{1}\hat{x}_{\lambda} = i\delta_{\lambda0}\mathbb{I} \,\, .
\end{equation}

If the Hamiltonian $\hat{H}$ is time-independent, 
\begin{equation}
\left[i\partial_{x_{0}},\hat{H}\right] = 0 \,\, ,
\label{time-independent-H}
\end{equation}
it depends on the momentum $\hat{P}_{1}$ in (\ref{generators}) and
$\hat{x}_{1}^{L}$, and we can write it as
\begin{equation}
\hat{H} = \hat{H}\left(\hat{x}_{1}^{L},\hat{P}_{1}\right) \,\, .
\end{equation}
It can depend on $\hat{x}_{1}^{R}$ as well if we rely just on
(\ref{time-independent-H}). But since $\hat{x}_{1}^{R}=-{\rm
  ad}\,\hat{x}_{1}+\hat{x}_{1}^{L}$, that means $\hat{H}$ has
dependence also on $i\partial_{x_{0}}$ and we can write
\begin{equation}
\hat{H} = 
\hat{H}\left(\hat{x}_{1}^{L},\hat{P}_{1},i\partial_{x_{0}}\right)\,\,.
\end{equation}
This generalisation however seems unwarranted: there is never such
dependence of $H$ on $i\partial_{x_{0}}$ for $\theta=0$, and we
will generally obtain $\hat{H}$ from $H$ in a manner that does not
induce this dependence.

If $\hat{H}$ has time-dependence and (\ref{time-independent-H})
is not correct, it will have $\hat{x}_{0}^{L}$, $\hat{x}_{0}^{R}$
or both in its arguments. But 
$\hat{x}_{0}^{L} = \theta \hat{P}_{1} + \hat{x}_{1}^{R}$, so in
the time-dependent case we write 
\begin{equation}
\hat{H} =
\hat{H}\left(\hat{x}_{0}^{R},\hat{x}_{1}^{L},\hat{P}_{1} \right)\,\,,
\label{x0R-dependence}
\end{equation}
ignoring a possible $i\partial_{x_{0}}$ dependence for reasons above.

The family of vector states constrained by the Schr\"{o}dinger equation
is
\begin{equation}
\tilde{\mathcal{H}}_{\theta}=\left\{
\hat{\psi}\in\mathcal{A}_{\theta} \left(\mathbb{R}^{2}\right):
\left(i\partial_{x_{0}}-\hat{H} \right) \hat{\psi} = 0 \right\}
\,\,,
\label{constrained_vectors}
\end{equation}
where arguments of $\hat{H}$ can be appropriately inserted.

The solutions of (\ref{constrained_vectors}) are easy to come by.
For the time-independent case,
\begin{equation}
\hat{\psi}\in\tilde{\mathcal{H}}_{\theta}\Longrightarrow\hat{\psi} =
e^{-i \left( \hat{x}_{0}^{R} - \tau_{I} \right)  
\hat{H}\left(\hat{P}_{1},\hat{x}_{1}^{L}\right)}
\hat{\chi}\left(\hat{x}_{1}\right) \,\, .
\label{formula_psi_1}
\end{equation}
The product $\hat{x}_{0}^{R}\hat{H}$ has no ordering problem since
$\left[\hat{x}_{0}^{R},\hat{H}\left(\hat{x}_{1}^{L},\hat{P}_{1}\right)\right]
=0$. Also $\tau_{I}$ is the initial time when $\hat{\psi} =
\hat{\chi}$. Since $\hat{x}_{0}^{R}$, $\hat{x}_{1}^{L}$ 
occur in the first factor, we should read the R.H.S. as the
exponential acting on the algebra element
$\hat{\chi}\left(\hat{x}_{1}\right)$. 

Suppose next that $\hat{H}$ depends on $\hat{x}_{0}^{R}$ as in
(\ref{x0R-dependence}). As $\hat{x}_{0}^{R}$ commutes with
$\hat{P}_{1}$ and $\hat{x}_{1}^{L}$, we can easily generalise the
formula (\ref{formula_psi_1}) to write
\[
\hat{\psi}\in\tilde{\mathcal{H}}_{\theta}\Longrightarrow\hat{\psi} =
U\left(\hat{x}_{0}^{R},\tau_{I}\right) 
\hat{\chi}\left(\hat{x}_{1}\right) \,\,,
\]
\begin{equation}
U\left(\hat{x}_{0}^{R},\tau_{I}\right) =
\left. T \exp\left[-i\left(\int_{\tau_{I}}^{x_{0}}
d\tau\,\hat{H}\left(\tau,\hat{x}_{1}^{L},\hat{P}_{1}
\right) \right) \right] \right|_{x_{0} =
\hat{x}_{0}^{R}} \,\, .
\label{formula_psi_2}
\end{equation}
Just as in (\ref{formula_psi_1}), the dependence of $U$ on $\hat{x}_{0}^{R}$
and $\tau_{I}$ has been displayed, while $\tau_{I}$ is the initial
time when $\hat{\psi}=\hat{\chi}$. 

Time translation by amount $\tau$ shifts $\hat{x}_{0}^{R}$ to
$\hat{x}_{0}^{R}+\tau$ in both (\ref{formula_psi_1}) and
(\ref{formula_psi_2}). 

An alternative useful form for $\hat{\psi}$ in (\ref{formula_psi_2}) is 
\begin{equation}
\hat{\psi} = V \left( \hat{x}_{0}^{R},-\infty \right) 
\hat{\chi} \left( \hat{x}_{1}  \right) \,\, ,
\end{equation}
\begin{equation}
V \left( \hat{x}_{0}^{R},-\infty \right) =
T \exp \left[ -i \int_{-\infty}^{0} d\tau \, \hat{H} 
\left( \hat{x}_{0}^{R}+\tau,\hat{x}_{1}^{L},\hat{P}_{1}\right)
\right] \,\, ,
\end{equation}
where the integral can be defined at the lower limit using the usual
adiabatic cut-off.

The Hilbert spaces $\mathcal{H}_{\theta}^{S}$ and $\mathcal{H}_{\theta}^{V}$
based on scalar products $\left(.,.\right)_{S}$ and $\left(.,.\right)_{V}$
are obtained from $\tilde{\mathcal{H}}_{\theta}$ by completion. Our
basic assumption is that $\hat{H}$ is self-adjoint in the chosen
scalar product. Then as before, the resultant Hilbert space
$\mathcal{H}_{\theta}^{S}$ or $\mathcal{H}_{\theta}^{V}$ has no
dependence on $t$.  

Assuming that
\begin{equation}
\hat{H} = \frac{\hat{P}_{1}^{2}}{2m}+V\left(\hat{x}_{1}\right)
\end{equation}
is a self-adjoint Hamiltonian for $\theta=0$, then we note that
its $\theta\ne0$ version
\begin{equation}
\hat{H} = \frac{\hat{P}_{1}^{2}}{2m}+V\left(\hat{x}_{1}^{L}\right)
\end{equation}
is self-adjoint on both $\mathcal{H}_{\theta}^{S}$ and
$\mathcal{H}_{\theta}^{V}$.  

If $\hat{H}\left(\hat{x}_{0},\hat{x}_{1},\hat{P}_{1}\right)$ is time-dependent
for $\theta=0$, we can form its $\theta\ne0$ version
\begin{equation}
\hat{H}\left(\hat{x}_{0}^{L},\hat{x}_{1}^{L},\hat{P}_{1}\right) =
\hat{H}\left( -\theta\hat{P}_{1} + \hat{x}_{0}^{R}, \hat{x}_{1}^{L},
\hat{P}_{1} \right) \,\, . 
\end{equation}
As $\hat{x}_{0}^{L}$ and $\hat{P}_{1}$ do not commute with $\hat{x}_{1}^{L}$,
we should check this $\hat{H}$ for factor-ordering problems. But
for this potential trouble, $\hat{H}$ is self-adjoint if $H$ is.

\subsubsection{Remarks on Time for $\theta\ne0$}

In the passage from $H$ to $\hat{H}$, there is an apparent ambiguity.
Above we replaced $x_{0}$ by $\hat{x}_{0}^{L}$, but we may be tempted
to replace $x_{0}$ by $\hat{x}_{0}^{R}$. In that case the passage
to $\theta\ne0$ will involve no factor-ordering problem as $\hat{x}_{0}^{R}$
commutes with $\hat{x}_{1}^{L}$ and $\hat{P}_{1}$. At the same time,
$\theta$-dependent terms in $\hat{H}$ disappear.

But it is incorrect to replace $x_{0}$ by $\hat{x}_{0}^{R}$ and
at the same time $x_{1}$ by $\hat{x}_{1}^{L}$. Time and space should
fulfill the relation (\ref{nc_relation}) when $\theta$ becomes nonzero
whereas $\hat{x}_{0}^{R}$ and $\hat{x}_{1}^{L}$ commute.

Note that $\hat{x}_{0}^{L,R}$ do not preserve the Schr\"{o}dinger constraint
so that there is no time operator for $\theta\ne0$ as well.

\subsubsection{Time-dependence for $\theta=0$ $\Longrightarrow$
Spatial nonlocality for $\theta\ne0$} 

We noted above that $\hat{x}_{0}^{L}=-\theta\hat{P}_{1}+\hat{x}_{0}^{R}$
and that $\hat{x}_{0}^{R}$ behaves much like the $\theta=0$ time
$x_{0}$. \textit{Thus if} $H$ \textit{has time-dependence, its effect
on $\hat{H}$ is to induce new momentum-dependent terms.} The
$x_{0}$-dependence in $H$ need not to be polynomial so that in $\hat{H}$
they induce nonpolynomial interactions in momentum, that is, instantaneous
spatially nonlocal (``acausal'') interactions.

\subsubsection{Observables}

We can construct observables as in (\ref{hat-K}) or its
version for time-dependent Hamiltonians. No complications are encountered.

\subsubsection{The Scalar Products $\left(.,.\right)_{S}$ and
$\left(.,.\right)_{V}$} 

We now explore the relation between the different scalar products.
We assume (as is often the case) that $\hat{H}$ is self-adjoint for
both. As the scalar products do not have time dependence, we have dropped
their time-subscripts.

It is enough to consider time-independent $\hat{H}$. Let $\hat{\psi}_{n}$
be its eigenstates,
\begin{equation}
\hat{H}\hat{\psi}_{n}=E_{n}\hat{\psi}_{n} \,\, ,
\end{equation}
and assume in the first instance that eigenvalues are non-degenerate:
\begin{equation}
E_{n} \ne E_{m} \,\, \textrm{if} \,\, n \ne m \,\, .
\end{equation}
For simplicity, the eigenvalues are taken to be discrete throughout
this discussion. Then since $\hat{H}$ is self-adjoint in either scalar
product, 
\begin{equation}
\left(\hat{\psi}_{m},\hat{\psi}_{n}\right)_{S} =
s_{n}\delta_{mn} \,\, , \,\,
\left(\hat{\psi}_{m},\hat{\psi}_{n}\right)_{V} =
v_{n}\delta_{mn}\,\,,\,\,
s_{n},v_{n}>0\,\,.
\end{equation}
Thus an isometry from the $S$-Hilbert space $\mathcal{H}_{\theta}^{S}$
to the $V$-Hilbert space $\mathcal{H}_{\theta}^{V}$ (based on the
scalar products $\left(.,.\right)_{S}$ and $\left(.,.\right)_{V}$
respectively) is 
\begin{equation}
\mathcal{H}_{\theta}^{S} \ni \frac{1}{\sqrt{s_{n}}} \hat{\psi}_{n}
\rightarrow
\frac{1}{\sqrt{v_{n}}}\hat{\psi}_{n} \in \mathcal{H}_{\theta}^{V} \,\, .
\end{equation}
If $\hat{K}_{S}$ is an observable in the $S$-Hilbert space with matrix
$k^{S}$ in the basis $\{ (1/\sqrt{s_{n}}) \hat{\psi}_{n} \}$,
\begin{equation}
\hat{K}_{S} \frac{1}{\sqrt{s_{n}}} \hat{\psi}_{n} =
\frac{1}{\sqrt{s_{m}}} \hat{\psi}_{m}k_{mn}^{S} \,\, ,\,\,
k_{mn}^{S}\in\mathbb{C}\,\, , 
\label{Ks-eigenvalues}
\end{equation}
to it we can associate the operator $\hat{K}_{V}$ on the $V$-Hilbert
space defined by
\begin{equation}
\hat{K}_{V} \frac{1}{\sqrt{v_{n}}} \hat{\psi}_{n} =
\frac{1}{\sqrt{v_{m}}} \hat{\psi}_{m}k_{mn}^{S}\,\, .
\label{Kv-eigenvalues}
\end{equation}
Then
\begin{equation}
\left( \frac{1}{\sqrt{s_{m}}} \hat{\psi}_{m},\hat{K}_{S}
\frac{1}{\sqrt{s_{n}}} \hat{\psi}_{n}\right)_{S} =
\left( \frac{1}{\sqrt{v_{m}}} \hat{\psi}_{m},\hat{K}_{V}
\frac{1}{\sqrt{v_{n}}} \hat{\psi}_{n}\right)_{V}
\end{equation}
and physics in the two spaces become identical.

If $\hat{H}$ has degeneracies, we can introduce a degeneracy index
$r$ and write
\[
\hat{H}\hat{\psi}_{n}^{(r)} =
E_{n}\hat{\psi}_{n}^{(r)} \,\, , \,\, E_{n\ } \ne E_{m} \,\,
\textrm{if} \,\, 
n\ne m\,\,,
\]
\begin{equation}
\left(\hat{\psi}_{m}^{(r)}, \hat{\psi}_{n}^{(s)}\right)_{S}=
s_{m}\delta_{rs}\delta_{mn}\,\,,\,\, s_{m}>0 \,\, .
\end{equation}
Then
\begin{equation}
\left(\hat{\psi}_{m}^{(r)},\hat{\psi}_{n}^{(s)}\right)_{V}=
\delta_{mn}W_{rs}(m) \,\, ,
\end{equation}
where $W(m)$ is a positive matrix with a positive invertible square
root $W(m)^{1/2}$. An isometry from $\mathcal{H}_{\theta}^{S}$
to $\mathcal{H}_{\theta}^{V}$ is thus 
\begin{equation}
\mathcal{H}_{\theta}^{S} \ni \frac{1}{\sqrt{s_{n}}}\hat{\psi}_{n}^{(r)} 
\rightarrow
\hat{\psi}_{n}^{(s)}W_{sr}(n)^{-1/2}
\in \mathcal{H}_{\theta}^{V} \,\, ,
\end{equation}
as is shown using $\left[ W(n)^{-1/2} \right]^{\dagger} = W(n)^{-1/2}$.

Following (\ref{Ks-eigenvalues}) and (\ref{Kv-eigenvalues}), we can also
map an observable $\hat{K}_{S}$ in $\mathcal{H}_{\theta}^{S}$ to
its equivalent on $\mathcal{H}_{\theta}^{V}$. Write
\begin{equation}
\hat{K}_{S} \frac{1}{\sqrt{s_{n}}} \hat{\psi}_{n}^{(r)} =
\frac{1}{\sqrt{s_{n'}}} \hat{\psi}_{n'}^{(r')}k_{n'r',nr}^{S}\,\, .
\end{equation}
Then
\begin{equation}
\hat{K}_{V}\hat{\psi}_{n}^{(s)} W_{sr}(n)^{-1/2} =
\hat{\psi}_{n'}^{(s)}W_{sr'}(n')^{-1/2}k_{n'r',nr}^{S}
\,\,. 
\end{equation}

Similar results are correct for time-dependent $\hat{H}$ and for
any scalar product compatible with the self-adjointness of $\hat{H}$.

We note that equivalent observables as elements of the algebra generally
differ for such differing scalar products. One universal exception
is the Hamiltonian when it is time-independent.

%%%%%%%%%%%%%%%%%%%%%%%%%%%%%%%%%%%%%%%%%%%%%%%%%%%%%%%%%%%%%%%%%%%%%%%%%%%%%
\section{Examples}
%%%%%%%%%%%%%%%%%%%%%%%%%%%%%%%%%%%%%%%%%%%%%%%%%%%%%%%%%%%%%%%%%%%%%%%%%%%%%

For definiteness, we work hereafter with $\mathcal{H}_{\theta}^{S}$.

\subsection{Plane Waves}

Let 
\begin{equation}
\hat{H}_{0} = \frac{\hat{P}_{1}^{2}}{2m}
\end{equation}
be the free Hamiltonian. Its eigenstates are 
\begin{equation}
\hat{\psi}_{k} =
e^{ik\hat{x}_{1}}e^{-i\omega(k) \hat{x}_{0}} \,\, , \,\,\omega(k) =
\frac{k^{2}}{2m} \,\, ,\,\,  k\in\mathbb{R} \,\, .
\end{equation}
The eigenvalues are $k^{2}/2m$: 
\begin{equation}
\hat{H}_{0}\hat{\psi}_{k} =
\left(\hat{H}_{0}e^{ik\hat{x}_{1}}\right)e^{-i\omega(k) \hat{x}_{0}} =
\omega(k) \hat{\psi}_{k} \,\,.
\end{equation}
The second factor in $\hat{\psi}_{k}$ is dictated by the Schr\"{o}dinger
constraint: 
\begin{equation}
\hat{P}_{0}\hat{\psi}_{k} =
e^{ik\hat{x}_{1}}\hat{P}_{0}e^{-i\omega(k) \hat{x}_{0}} =
\omega(k) \hat{\psi}_{k} \Longrightarrow
\left(\hat{P}_{0}-\hat{H}\right)\hat{\psi}_{k} = 0 \,\, .
\end{equation}
The spectrum of $\hat{H}_{0}$ is completely conventional while the
noncommutative plane waves too resemble the ordinary plane waves.
But phenomena like beats and interference show new features
\cite{Balachandran}. 

The coincidence of spectra of the free Hamiltonians in commutative
and noncommutative cases is an illustration of a more general result
which we now establish.

\subsection{A Spectral Map}

For $\theta=0$ consider the Hamiltonian
\begin{equation}
H = -\frac{1}{2m}\frac{\partial^{2}}{\partial x_{1}^{2}}+V(\hat{x}_{1})
\end{equation}
with eigenstates $\psi_{E}$ fulfilling the Schr\"{o}dinger constraint:
\begin{equation}
\psi_{E}\left(\hat{x}_{0},\hat{x}_{1}\right) =
\varphi_{E}(\hat{x}_{1})e^{-iE\hat{x}_{0}} \,\, ,
\end{equation}
\begin{equation}
H\varphi_{E} = E\varphi_{E}\,\,.\
\label{hamilt_eigens}
\end{equation}

The Hamiltonian $\hat{H}$ associated to $H$ for $\theta\ne0$ is
\begin{equation}
\hat{H} = \frac{\hat{P}_{1}^{2}}{2m}+V(\hat{x}_{1}) \,\, .
\label{nc_free_H}
\end{equation}
Then $\hat{H}$ has exactly the same spectrum as $H$ while its eigenstates
$\hat{\psi}_{E}$ are obtained from $\psi_{E}$ just by regarding
$\hat{x}_{0}$ and $\hat{x}_{1}$ as fulfilling (\ref{nc_relation}):
\begin{equation}
\hat{\psi}_{E}=\varphi_{E}(\hat{x}_{1})e^{-iE\hat{x}_{0}} \,\, ,
\end{equation}
\begin{equation}
\hat{H}\varphi_{E}(\hat{x}_{1})=E\varphi_{E}(\hat{x}_{1}) \,\, .
\label{eingen-ind-time}
\end{equation}

The proof of (\ref{eingen-ind-time}) follows from (\ref{nc_free_H})
as it involves no feature associated with spacetime noncommutativity.
Since
\begin{equation}
\hat{P}_{0}\hat{\psi}_{E} =
\varphi_{E}(\hat{x}_{1})\hat{P}_{0}e^{-iE\hat{x}_{0}} =
E\hat{\psi}_{E} \,\, ,
\end{equation}
we see that $\hat{\psi}_{E}$ fulfills the Schr\"{o}dinger constraint
as well.

When the spatial slice for a commutative spacetime $\mathbb{R}^{d}$ is
of dimension two or larger, one can introduce space-space noncommutativity as
well. That would change the noncommutative Hamiltonian. The spectral
map may not then exist.

%%%%%%%%%%%%%%%%%%%%%%%%%%%%%%%%%%%%%%%%%%%%%%%%%%%%%%%%%%%%%%%%%%%%%%%%%%%%%
\section{Conserved Current }
%%%%%%%%%%%%%%%%%%%%%%%%%%%%%%%%%%%%%%%%%%%%%%%%%%%%%%%%%%%%%%%%%%%%%%%%%%%%%

The existence of a current $j_{\lambda}$ which fulfills
the continuity equation has a particular importance when $\theta=0$.
It is this current which after second quantization couples to
electromagnetism \cite{Pinzul}.    

There is such a conserved current also for $\theta\ne0$. It follows
in the usual way from (\ref{constrained_vectors}) and its $*$: 
\begin{equation}
\left( \hat{P}_{0} \hat{\psi} \right)^{*} - \hat{\psi}^{*} \hat{H}
= - \hat{P}_{0} \hat{\psi}^{*} -  \hat{\psi}^{*} \hat{H}
= 0 \,\, .
\label{star_schrodinger_equation}
\end{equation}
Here we assumed that $\hat{V}^{*}=\hat{V}$.

Multiplying the Schr\"{o}dinger constraint in (\ref{constrained_vectors})
on left by $\hat{\psi}^{*}$ and (\ref{star_schrodinger_equation})
on right by $\hat{\psi}$ and subtracting, 
\begin{equation}
\hat{P}_{0}\left(\hat{\psi}^{*}\hat{\psi}\right) =
\hat{\psi}^{*}\left(\frac{\hat{P}_{1}^{2}}{2m} \hat{\psi} \right)
- \left(\frac{\hat{P}_{1}^{2}}{2m}\hat{\psi}^{*} \right) \hat{\psi}
= \frac{\hat{P}_{1}}{2m}\left[
\hat{\psi}^{*}\left(\hat{P}_{1}\hat{\psi}\right)
- \left(\hat{P}_{1}\hat{\psi}^{*}\right)\hat{\psi}
\right] \,\,.
\label{continuity_equation}
\end{equation}
With
\begin{equation}
\hat{\rho} = \hat{\psi}^{*}\hat{\psi}
\,\,,\,\,
\hat{j} =  \frac{1}{2m}
\left[\hat{\psi}^{*}\left(\hat{P}_{1}\hat{\psi}\right) -
\left(\hat{P}_{1}\hat{\psi}^{*}\right)\hat{\psi}\right]
\end{equation}
as the noncommutative charge and current densities, (\ref{continuity_equation})
can be interpreted as the noncommutative continuity equation.

%%%%%%%%%%%%%%%%%%%%%%%%%%%%%%%%%%%%%%%%%%%%%%%%%%%%%%%%%%%%%%%%%%%%%%%%%%%%%
\section{Towards Quantum Field Theory}
%%%%%%%%%%%%%%%%%%%%%%%%%%%%%%%%%%%%%%%%%%%%%%%%%%%%%%%%%%%%%%%%%%%%%%%%%%%%%

Perturbative quantum field theories (qft's) based on algebras like
(\ref{nc_relation}) have been treated with depth by Doplicher et
al. \cite{Doplicher}. We can also see how to do perturbative qft's, our
approach can be inferred from the work of Doplicher et al. 

In the interaction representation, an operator $U_{I}$ determines
the $S$-matrix. It is in turn determined by the interaction Hamiltonian
 $\hat{H}_{I}$. The latter is based on ``free fields''  which
are solutions of the Klein-Gordon equation (We assume zero spin for
simplicity). Examples of $\hat{H}_{I}$ can be based on interaction
Hamiltonians $H_{I}$ such as  $\lambda\int dx_{1}\Phi(x_{0},x_{1})^{4}$
 (with $\Phi^{\dagger}=\Phi$ being a free field) for $\theta=0$.
For this particular $H_{I}$, $\hat{H}_{I}$ can be something like
 $\lambda
S_{x_{0}}\left[\hat{\Phi}(\hat{x}_{0},\hat{x}_{1})^{4}\right]$
(cf. (\ref{S-positive-map})), where $\hat{\Phi}$ is the self-adjoint
free field for $\theta\ne0$. We make this expression more precise
below.  

We require of $\hat{\Phi}$ that it is a solution of the massive Klein-Gordon
equation:  
\begin{equation}
\left(\textrm{ad}\hat{P}_{0}^{2} -
\textrm{ad}\hat{P}_{1}^{2}+\mu^{2}\right)\hat{\Phi} = 0 \,\, .
\label{nc_KleinGordon_eq}
\end{equation}
The plane wave solutions of (\ref{nc_KleinGordon_eq}) are  
\begin{equation}
\hat{\phi}_{k} =
e^{ik\hat{x}_{1}}e^{-i\omega(k)\hat{x}_{0}} \,\, , 
\,\,\omega(k)^{2}-k^{2} = \mu^{2} \,\, .
\end{equation}
So for $\hat{\Phi}$, we write \cite{Chaichian} 
\begin{equation}
\hat{\Phi} =
\int\frac{dk}{2\omega(k)}\left[a_{k}\hat{\phi}_{k}+a_{k}^{\dagger}
\hat{\phi}_{k}^{\dagger}\right] \,\, , 
\label{free_field}
\end{equation}
where $a_{k}$ and $a_{k}^{\dagger}$ commute with $\hat{x}_{\mu}$
and define harmonic oscillators:
$\left[a_{k},a_{k}^{\dagger}\right]=2\omega(k)\delta(k-k')$.  

The expression (\ref{free_field}) is the ``free''  field ``coinciding
with the Heisenberg field initially''. After time translation by
amount $\tau$ using the free Schr\"{o}dinger Hamiltonian 
\begin{equation}
\hat{H}_{0} = \int\frac{dk}{2\omega(k)}a_{k}^{\dagger}a_{k}\,\,,
\end{equation}
it becomes 
\begin{equation}
U_{0}(\tau)\left(\hat{\Phi}\right) =
e^{i\tau\hat{H}_{0}}\hat{\Phi}e^{-i\tau\hat{H}_{0}} \,\, ,
\end{equation}
The interaction Hamiltonian is accordingly 
\begin{equation}
\hat{H}_{I}\left(x_{0}\right) =
\lambda:S_{x_{0}}\left(U_{0}(\tau)\left(\hat{\Phi}\right)^{4}\right):
\; = \lambda:S_{x_{0}+\tau}\left(\hat{\Phi}^{4}\right):\,\,,\lambda>0\,\,,
\end{equation}
where $:\;\;:$ denotes the normal ordering of $a_{k}$ and
$a_{k}^{\dagger}$. 
 
The $S$-matrix $S$ can be worked out as usual: 
\begin{equation}
S =
T \, \exp\left[-i\int_{-\infty}^{+\infty} d\tau \,
\lambda:S_{\tau}\left(\hat{\Phi}^{4}\right):\right] \,\, .
\label{S-matrix}
\end{equation}
It is important to recognise, as is clear from Doplicher et
al. \cite{Doplicher}, that time-ordering is with respect to the
time-translation parameter $\tau$ and not the spectrum of the operator
$\hat{x}_{0}^{L}$. Its perturbation series can be developed since we
understand the relevant properties of $\hat{\Phi}$. 

Scattering amplitudes can be calculated from (\ref{S-matrix}). There
is no obvious reason why they are not compatible with perturbative
unitarity \cite{Bak}.

\acknowledgments

We have benefitted from discussions with K. S. Gupta, P. Presnajder,
A. Pinzul, A. Stern and S. Vaidya. Kumar Gupta and Peter Presnajder
were particularly helpful with detailed comments. We also thank
W. Bietenholz for the reference to Yang in \cite{Yang} and  R. Jackiw for
reminding us of the work in \cite{Susskind,Jackiw}.
This work was partially supported by DOE under contract number
DE-FG02-85ER40231, by NSF under contract number INT9908763 and by
FAPESP, Brazil.

%%%%%%%%%%%%%%%%%%%%%%%%%%%%%%%%%%%%%%%%%%%%%%%%%%%%%%%%%%%%%%%%%%%%%%%%%%%%%

\end{document}